\documentclass[conference]{IEEEtran}
\usepackage{cite}
\usepackage{amsmath,amssymb,amsfonts}
\usepackage{algorithmic}
\usepackage{graphicx}
\usepackage{textcomp}
\usepackage{xcolor}
\usepackage{wrapfig}
\usepackage{comment}
\usepackage{amsthm}
\usepackage{longtable}
\usepackage{ upgreek }
\usepackage{amsmath}
\usepackage{graphicx} 
\usepackage{ gensymb }
\usepackage{ dsfont }
\usepackage{tabularx,booktabs}
\usepackage[acronym,nomain,nonumberlist]{glossaries}
\usepackage{glossary-tree}
\usepackage{tabularx}
\newcolumntype{L}{>{\raggedright\arraybackslash}X}

\usepackage{cleveref}
\usepackage{subcaption}
\usepackage{makecell}
\usepackage{fancyhdr}

\usepackage[ruled,norelsize]{algorithm2e}
\makeatletter
\newcommand{\removelatexerror}{\let\@latex@error\@gobble}
\makeatother
\SetKw{And}{\hspace{\algoskipindent}\itshape and\;}
\SetKwBlock{Condition}{}{}
\SetKw{Or}{\hspace{\algoskipindent}\itshape or\;}

\makeglossaries
\newacronym{5g}{5G}{Fifth Generation}
\newacronym{mmtc}{mMTC}{massive Machine-Type Com\-mu\-ni\-ca\-tions}
\newacronym{urllc}{URLLC}{Ultra-Reliable Low Latency Communications}
\newacronym{embb}{eMBB}{enhanced Mobile Broadband}
\newacronym{tn}{TN}{Terrestrial Network}
\newacronym{dynasat}{DYNASAT}{"Dynamic spectrum sharing and bandwidth-efficient techniques for high-through\-put MIMO Satellite systems"}
\newacronym{ntn}{NTN}{Non-Terrestrial Network}
\newacronym{mc}{MC}{Multi-Connectivity}
\newacronym{iot}{IoT}{Internet of Things}
\newacronym{vr}{VR}{Virtual Reality}
\newacronym{ngso}{NGSO}{Non-Geostationary Orbit}
\newacronym{dsa}{DSA}{Dynamic Spectrum Allocation}
\newacronym{poc}{PoC}{Proof-of-Concept}
\newacronym{3gpp}{3GPP}{3rd Generation Partnership Project}
\newacronym{ue}{UE}{User Equipment}
\newacronym{dc}{DC}{Dual Connectivity}
\newacronym{mn}{MN}{Master Node}
\newacronym{sn}{SN}{Secondary Node}
\newacronym{mrdc}{MR-DC}{Multi Radio-Dual Connectivity}
\newacronym{eutra}{E-UTRA}{Evolved Universal Terrestrial Access}
\newacronym{enb}{eNB}{Evolved Node B}
\newacronym{epc}{EPC}{Evolved Packet Core}
\newacronym{5gc}{5GC}{5G Core}
\newacronym{nr}{NR}{New Radio}
\newacronym{engnb}{en-gNB}{en-Next Generation Node B}
\newacronym{ngenb}{ng-eNB}{Next Generation eNB}
\newacronym{ca}{CA}{Carrier Aggregation}
\newacronym{cc}{CC}{Component Carrier}
\newacronym{rrm}{RRM}{Radio Resource Management}
\newacronym{pdcp}{PDCP}{Packet Data Convergence Protocol}
\newacronym{mac}{MAC}{Media Access Control}
\newacronym{rrc}{RRC}{Radio Resource Control}
\newacronym{cn}{CN}{Core Network}
\newacronym{mcg}{MCG}{Master Cell Group}
\newacronym{scg}{SCG}{Secondary Cell Group}
\newacronym{rf}{RF}{Radio Frequency}
\newacronym{ngran}{NG-RAN}{Next Generation Radio Access Network}
\newacronym{gnbcu}{gNB-CU}{gNB-Centralized Unit}
\newacronym{gnbdu}{gNB-DU}{gNB-Distributed Unit}
\newacronym{leo}{LEO}{Low Earth Orbit}
\newacronym{geo}{GEO}{Geostationary Orbit}
\newacronym{ran}{RAN}{Radio Access Network}
\newacronym{hetnet}{HetNet}{Heterogeneous Networks}
\newacronym{rsrp}{RSRP}{Reference Signal Received Power}
\newacronym{ns3}{ns-3}{Network Simulator 3}
\newacronym{gnb}{gNB}{Next Generation Node B}
\newacronym{e2e}{E2E}{End-to-End}
\newacronym{pgw}{PGW}{Packet Network Data Gateway}
\newacronym{sgw}{SGW}{Serving Gateway}
\newacronym{amf}{AMF}{Access and Mobility Management Function}
\newacronym{upf}{UPF}{User Plane Function}
\newacronym{ra}{RA}{Random Access}
\newacronym{4g}{4G}{Fourth Generation}
\newacronym{ap}{AP}{Access Point}
\newacronym{srs}{SRS}{Sounding Reference Signal}
\newacronym{udp}{UDP}{User Datagram Protocol}
\newacronym{sls}{SLS}{System Level Simulator}
\newacronym{kpi}{KPI}{Key Performance Indicator}
\newacronym{ecdf}{eCDF}{empirical Cumulative Distribution Function}
\newacronym{tcp}{TCP}{Transmission Control Protocol}
\newacronym{ahp}{AHP}{Analytic Hierarchy Process}
\newacronym{rat}{RAT}{Radio Access Technology}
\newacronym{sinr}{SINR}{Signal-to-Interference-plus-Noise Ratio}
\newacronym{ts}{TS}{Technical Specification}
\newacronym{tr}{TR}{Technical Report}
\newacronym{lan}{LAN}{Local Area Network}
\newacronym{rnti}{RNTI}{Radio Network Temporary Identifier}
\newacronym{sib}{SIB}{System Information Block}
\newacronym{mib}{MIB}{Master Information Block}
\newacronym{nlos}{NLOS}{Non-Line of Sight}
\newacronym{rng}{RNG}{Random Number Generator}
\newacronym{sue}{SUE}{Spectral Utilization Efficiency}
\newacronym{rb}{RB}{Resource Block}
\newacronym{re}{RE}{Resource Element}
\newacronym{ewma}{EWMA}{Exponential Weighted Moving Average}
\newacronym{sdap}{SDAP}{Service Data Adaption Protocol}
\newacronym{ho}{HO}{Handover}
\newacronym{pdu}{PDU}{Protocol Data Unit}
\newacronym{wa}{WA}{Wraparound}
\newacronym{cbr}{CBR}{Constant Bit Rate}

\def\BibTeX{{\rm B\kern-.05em{\sc i\kern-.025em b}\kern-.08em
    T\kern-.1667em\lower.7ex\hbox{E}\kern-.125emX}}

\renewcommand\IEEEkeywordsname{Keywords}

\makeatletter
\newcommand{\linebreakand}{%
  \end{@IEEEauthorhalign}
  \hfill\mbox{}\par
  \mbox{}\hfill\begin{@IEEEauthorhalign}
}
\makeatother

\fancypagestyle{FirstPage}{

\lfoot{\copyright 2023 IEEE. Personal use of this material is permitted. Permission from IEEE must be obtained for all other uses, in any current or future media, including reprinting/republishing this material for advertising or promotional purposes, creating new collective works, for resale or redistribution to servers or lists, or reuse of any copyrighted component of this work in other works. DOI 10.1109/WiSEE58383.2023.10289158}

}

\begin{document}
\bstctlcite{IEEEexample:BSTcontrol}
\newtheorem{thm}{Theorem} 
\theoremstyle{definition}
\newtheorem{remark}[thm]{Remark}
\newtheorem{defn}[thm]{Definition}
\theoremstyle{plain}
\newtheorem{thr}[thm]{Theorem}
\newtheorem{prop}[thm]{Proposition}
\newtheorem{kor}[thm]{Corollary}

\title{On Enhancing Reliability in B5G NTNs with Packet Duplication via Multi-Connectivity}

\author{
\IEEEauthorblockN{Mikko Majamaa\IEEEauthorrefmark{1}\IEEEauthorrefmark{2}, Henrik Martikainen\IEEEauthorrefmark{1}, Jani Puttonen\IEEEauthorrefmark{1} and Timo Hämäläinen\IEEEauthorrefmark{2}}

\IEEEauthorblockA{
\IEEEauthorrefmark{1}\textit{Magister Solutions, Jyv\"{a}skyl\"{a}, Finland} \\
email: \{firstname.lastname\}@magister.fi
}

\IEEEauthorblockA{
\IEEEauthorrefmark{2}\textit{Faculty of Information Technology, University of Jyv\"{a}skyl\"{a}, Jyv\"{a}skyl\"{a}, Finland} \\
email: timo.t.hamalainen@jyu.fi}

}

\maketitle

\begin{abstract}

Non-Terrestrial Networks (NTNs) can be used to provide ubiquitous 5G and beyond services to un(der)served areas. To ensure reliable communication in such networks, packet duplication (PD) through multi-connectivity is a promising solution. However, the existing PD schemes developed for terrestrial environments may not be reactive enough for the NTN environment where propagation delays are significantly longer. This paper proposes a dynamic PD activation scheme for NTNs based on hybrid automatic repeat request feedback. The scheme aims to reduce the number of duplicated packets while maintaining high reliability. To evaluate the proposed scheme, simulations are conducted in a scenario with two transparent payload low-earth orbit satellites. The results show a significant reduction of 87.2\% in the number of duplicated packets compared to blind duplication, with only marginal compromise in reliability.

\end{abstract}


\section{Introduction}
\label{sec:introduction}

\thispagestyle{FirstPage}

Non-Terrestrial Networks (NTNs) can be used to provide ubiquitous 5G and beyond services to un(der)served areas. 3GPP Release-17 \mbox{(Rel-17)} included basic functionalities to provide New Radio (NR), the air interface of 5G, access via NTNs. \mbox{Rel-18} will enhance the functionalities by improving coverage for handheld devices, considering deployment to higher frequencies, and addressing mobility aspects \cite{wid}.

A recent report by ITU \cite{itu} defines the requirements for satellite radio interfaces of IMT-2020. Whereas the terrestrial 5G has the Ultra-Reliable Low Latency Communications (URLLC) service class with a latency requirement of 1 ms and 99.999\% reliability, the satellite equivalent service class is called High-Reliability Communications via satellite \mbox{(HRC-s)} with a reliability requirement of 99.9\%. To achieve higher reliability, Packet Duplication (PD) is a promising solution.

PD can be achieved through Multi-Connectivity (MC). In MC, a User Equipment (UE) can be connected to multiple base stations simultaneously. New Radio Dual Connectivity (NR-DC) \cite{37340} is a form of MC in which MC is achieved at the Packet Data Convergence Protocol (PDCP) layer. In NR-DC, a UE is connected to a Master Node (MN) and Secondary Node (SN) which both provide NR access. MC for TNs has been standardized by 3GPP. This is not the case for NTNs, but MC is one of the \mbox{Rel-19} candidates.

In the related literature, PD is typically performed either blindly (i.e., all the packets are duplicated) to all the users who have established a secondary connection, or the trigger to activate PD is based on a measure that is not reactive enough. The first issue is problematic because of the increased use of network resources. The latter may not be a problem in the TN environment where the propagation paths are relatively short. To this end, there is a need to research methods to quickly and dynamically react to the need to activate PD in the NTN environment. In this paper, a dynamic PD scheme is introduced in which PD is started as a precaution based on negative feedback from Hybrid Automatic Repeat reQuest (HARQ).

The paper is organized as follows. In the next section, related work is reviewed. In Section \ref{sec:packetduplication}, PD and the proposed PD scheme are discussed. The proposed scheme is evaluated through system simulations in Section~\ref{sec:simulations}. The paper is concluded in Section~\ref{sec:conclusions}.

\section{Related Work}
\label{sec:relatedwork}

Reference \cite{8884124} highlights MC's role as an enabler for reliable low latency communications, concluding that PD, along with load balancing and packet splitting, is one of the fundamental scheduling categories for meeting the requirements of such communications. The authors in \cite{8491093} study Dual Connectivity (DC) for reliability enhancement through PD in TNs. For the SN addition, the A3 event is used, that is, the neighbor cell Reference Signal Received Power (RSRP) becomes a threshold value less than the MN RSRP. Further, a DC range parameter is introduced that corresponds to a negative threshold for SN addition. Through numerical simulations in small and macro cells with different DC range values and Block Error Rate (BLER) targets, it is concluded that DC is a promising technique for reliability enhancement, particularly in scenarios with relaxed latency requirements. In \cite{9027289}, DC is investigated for improved reliability by dynamically duplicating packets when link quality drops below a threshold and signaling UE to use two links in the uplink direction, resulting in more efficient resource utilization compared to blind duplication in scenarios with both 4G and 5G access. Reference \cite{8377054} presents a detailed overview of MC architecture and introduces a dynamic algorithm that activates PD based on Signal-to-Noise Ratios (SNRs), transport block sizes, and reliability targets, leading to reduced resource usage, though with marginal savings in high SNR regimes and small packet sizes.

In \cite{9128536}, two methods to enhance the performance of PD in TNs are proposed. The first involves the UE notifying the SN when a packet is successfully received, reducing the need for duplicate packet transmissions. However, this approach may not be sufficiently reactive in NTN environments due to long propagation delays. The second method is based on HARQ feedback, where the MN performs PD upon receiving a NACK as HARQ feedback. This method is similar to the one introduced in this paper, but instead of duplicating a single packet after a NACK, we propose starting a timer during which duplication is performed. This allows to better capture consecutive NACKs and adapt to the long propagation delays.

It is worth noting that MC with PD for reliability enhancement has not been extensively researched in NTNs. In \cite{10008752}, PD was experimentally tested utilizing 5G TN and a connection to a Starlink Low Earth Orbit (LEO) satellite. The receiver used was a special type of antenna. Neither has research on MC in NTNs been extensively performed. The authors have previously investigated MC for throughput enhancement, for example, in \cite{majamaa2022multiconnectivity} in which algorithms for MC activation and traffic splitting are introduced. Further research is needed to explore the potential of MC with PD for reliability enhancement in NTN scenarios where the existing solutions for TNs are not sufficient due to the longer propagation delays.

\section{Packet Duplication}
\label{sec:packetduplication}
\subsection{General}
\label{sec:general}

\begin{figure}[htb!]
    \centering
    \includegraphics[width=.8\linewidth]{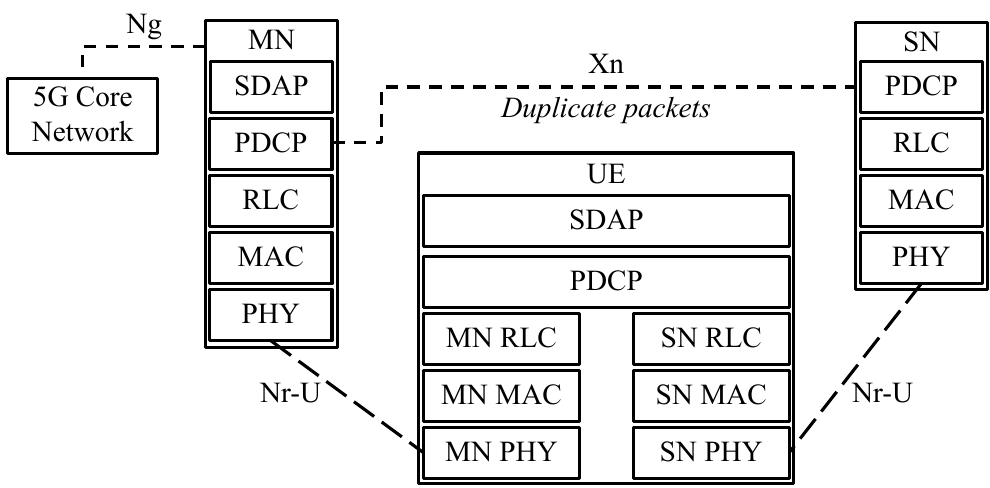}
    \caption{PD through MC illustrated. The MN and SN are connected through the Xn interface. The MN is connected to the core network through the Ng interface. Both the gNBs provide NR access to the UE through the Nr-U interface.}
    \label{fig:pd}
\end{figure}

In MC, a UE can be connected to multiple base stations simultaneously. NR-DC is a form of MC in which MC is achieved at the PDCP layer. In NR-DC, a UE is connected to a MN and SN which both provide NR access. The SN addition process is initiated by the current serving Next Generation Node B (gNB) of the UE. Based on some trigger, typically RSRP measurement received from the UE, it sends a request to a candidate SN.  The MN and SN can exchange user and control plane data through the Xn interface. 
When a packet arrives at the PDCP layer of the MN, it can duplicate the packet, and send a copy of it to the SN. In this way, the data is transmitted through both the MN and SN to increase the probability of reception of the data at the UE side. PD is illustrated in Fig.~\ref{fig:pd}. The UE hosts separate protocol stacks for MN and SN transmissions and combines them at the PDCP layer. The PDCP layer also detects and drops packets already received.

{}

\subsection{Dynamic Packet Duplication}

After activating MC to a UE, the MN can start sending duplicate packets for the SN to send to the UE. This can be done blindly, that is, all the packets are duplicated. However, this significantly increases the utilization of network resources, thus necessitating the exploration of more intelligent methods to activate PD. In this work, a solution to achieve this based on HARQ feedback is proposed.

In HARQ, the UE sends feedback based on the data it receives (ACK) or doesn’t (NACK). The UE knows to expect data from control channel signals. Commonly, the control channel is more robust than the data channel, so the control messages are less likely to get lost. In this work, PD is activated for a parametrizable period after a NACK is received as HARQ feedback in the primary connection. When a NACK is received, the gNB (i.e., the MN) starts a duplication timer for the corresponding UE. If the timer is running when a packet arrives at the MN's PDCP, it is duplicated. Each NACK resets and restarts the timer.

Note that, after receiving a NACK as HARQ feedback, the gNB can retransmit data to the UE, which may be successfully received. However, PD activation serves as a precaution against possible successive failures due to factors like Non-Line-of-Sight (NLOS) conditions. Additionally, data already forwarded from the PDCP layer may be lost, but this could be mitigated by maintaining a buffer of packet copies at the PDCP that can be forwarded to the SN in case a NACK is received as HARQ feedback.

\subsection{Delivery Options and Reordering in PDCP}

\begin{figure}[htb!]
    \centering
    \includegraphics[trim={0 0 0 0},clip, width=\linewidth]{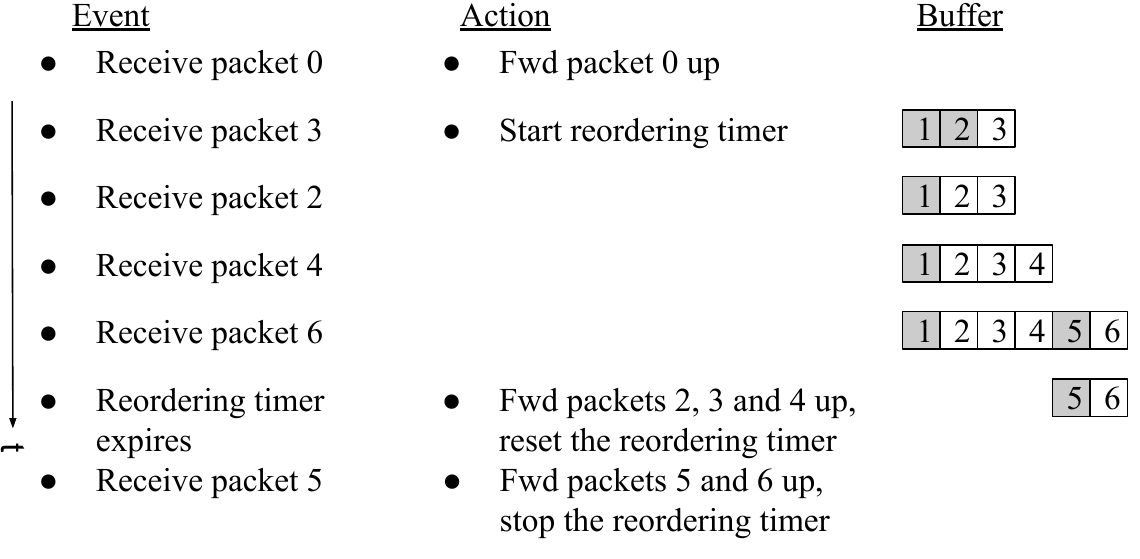}
    \caption{Illustration of in-order delivery at PDCP.}
    \label{fig:pdcp}
\end{figure}

The receiving PDCP performs duplicate detection, deciphering, and integrity verification of received packets \cite{38323}. If any of these operations fail, the packet is dropped, which is indicated to the upper layer. If the packet is not dropped, PDCP chooses between two delivery options: i) out-of-order delivery and ii) in-order delivery with reordering. The first one is the simpler mode in which the received packets are delivered to the upper layer without reordering. In the second option, PDCP tries to deliver packets in-order to the upper layer based on the sequence numbers of the packets. When out-of-order packets are received, a timer is started. If the missing packets are received when the timer is running, the packets are delivered and the timer is stopped. If the timer expires, the packets with sequence numbers below the one that caused the timer to start are delivered. In addition, the consecutively associated packets (according to their sequence numbers) starting from the one that caused the timer to start are delivered. If all the packets in the buffer are still not forwarded, the timer is reset and started again. In-order delivery at PDCP is illustrated in Fig~\ref{fig:pdcp}.

{}

In the figure, first, packet 0 is received from the lower layer and delivered up. After that, packet 3 is received from the lower layer, that is, packets 1 and 2 are missing (illustrated as grey color in the figure). This event triggers the reordering timer. While the timer is running, packets 2, 4, and 6 are received, but not the still-missing packet 1. Eventually, the reordering timer expires, and packets 2, 3, and 4 are delivered to the upper layer (if packet 1 is received from now on, it will be discarded because it is not expected anymore). Packet 6 cannot be forwarded, since packet 5 is missing so the reordering timer is reset. Then packet 5 arrives, packets 5 and 6 are delivered to the upper layer and the timer is stopped. 

The reordering window's length is configured by upper layers \cite{38331} and can vary from 0 ms (out-of-order delivery) to 3 s, or even set to infinity for forced in-order delivery. In NTN environments, the propagation delay differences between paths in MC can be significantly higher than in TN environments. However, the currently defined values for the reordering window's length should still be able to handle propagation delay differences, even in Geostationary Orbit (GEO) satellite cases. Although, an increase in UE's buffering requirement is expected in MC involving NTN compared to TN MC when in-order delivery is required. Further, if the propagation delays between the paths differ, the packets from the faster path require extra buffering in the receiver's PDCP before delivering up, thus increasing the packet delay of these packets as well. This needs to be considered in delay-sensitive applications.

\section{Evaluation}
\label{sec:simulations}

\subsection{5G Non-Terrestrial Network Simulator}

The simulator used for the assessment is a 5G NTN System-Level Simulator (SLS) \cite{ntn}, which is built on top of Network Simulator 3 (ns-3) \cite{Riley2010} and its 5G LENA module \cite{Patriciello2019AnES}. ns-3 is typically used for educational and research purposes, and users can add new modules to the simulator. The 5G LENA module is designed to simulate 5G networks, but it cannot simulate NTNs. Therefore, the necessary components to simulate NTNs were implemented in the 5G NTN SLS using 5G LENA as a starting point.

The 5G LENA module implements NR Physical (PHY) and Medium Access Control (MAC) features, while the upper layers of the UE/gNB stack are reused from the ns-3 Long Term Evolution (LTE) module \cite{lenalte}. The ns-3 core provides the higher layers such as transport and network layers, while the link layer is abstracted with Link-to-System (L2S) mapper and Modulation and Coding (MODCOD)-specific SINR to BLER mapping curves. SINR is computed for each packet, and using the mapper, BLER is deduced.

The simulator has been calibrated in the past using system-level calibration scenarios from 3GPP~Technical~Report~(TR)~38.821 \cite{38821}. The calibration scenarios provide different assumptions, such as bands (S-band/Ka-band), terminal types (VSAT, handheld), and frequency reuse patterns (reuse 1, 3, 2+2), and they can be adjusted as needed. Additionally, hybrid TN-NTN scenarios can be studied. Channel and antenna/beam modeling from TR~38.811 \cite{38811} have also been implemented in the simulator. Channel modeling is elaborated on in the following subsection. The MC feature in the simulator was implemented following the specifications outlined in 3GPP~Technical~Specification~(TS)~37.340 \cite{37340}.

\subsection{Channel Model}

The considered NTN channel model is based on TR 38.811. In the NTN environment, several attenuation factors affect the propagation of the signal. The total path loss (in dBs) is defined as

\begin{equation}
    PL = PL_b + PL_g + PL_s,
\end{equation}

\noindent
where $PL_b$ is the basic path loss, $PL_g$ is the attenuation due to atmospheric gasses, and $PL_s$ is the attenuation due to scintillation. The basic path loss in dB units is modeled as

\begin{equation}
    PL_b = FSPL(d,f_c) + SF + CL(\alpha,f_c),
\end{equation}

\noindent
where $FSPL(d,f_c)$ is the Free Space Path Loss, $d$ is the slant range between the satellite and UE, $f_c$ is the carrier frequency, $SF$~(${\sim} N(0,\sigma^2_{SF}$)) is the Shadow Fading loss, $CL(\alpha,f_c)$ is the Clutter Loss, and $\alpha$ is the elevation angle. CL refers to the reduction in signal power due to the presence of nearby buildings and objects on the ground, which attenuates the signal. CL is negligible when the UE is in Line-of-Sight (LOS) conditions. The values of $CL$ and $\sigma^2_{SF}$ can be found in Tables~6.6.2-1 to 6.6.2-3 in TR~38.811 for different scenarios and elevation angles. 

The reduction in signal strength caused by the absorption of atmospheric gases is primarily influenced by frequency, elevation angle, altitude above sea level, and water vapor density (absolute humidity). Typically, at frequencies below 10~GHz, this attenuation can be disregarded. However, it is advisable to consider this calculation for frequencies above 1~GHz when dealing with elevation angles below 10 degrees. Further, rain and cloud attenuation are considered negligible for frequencies below 6~GHz.

Now, the received signal power of a user can be computed as

\begin{equation}
    C = EIRP + G_\textnormal{Rx} - PL,
\end{equation}

\noindent
where EIRP is the Effective Isotropic Radiated Power from the satellite toward the user and  $G_\textnormal{Rx}$ is the receiver antenna gain. For more detailed link budget analysis, refer to \cite{9149179}.

\subsection{Scenario and Assumptions}

\begin{figure}[htb!]
    \centering
    \includegraphics[width=.8\linewidth]{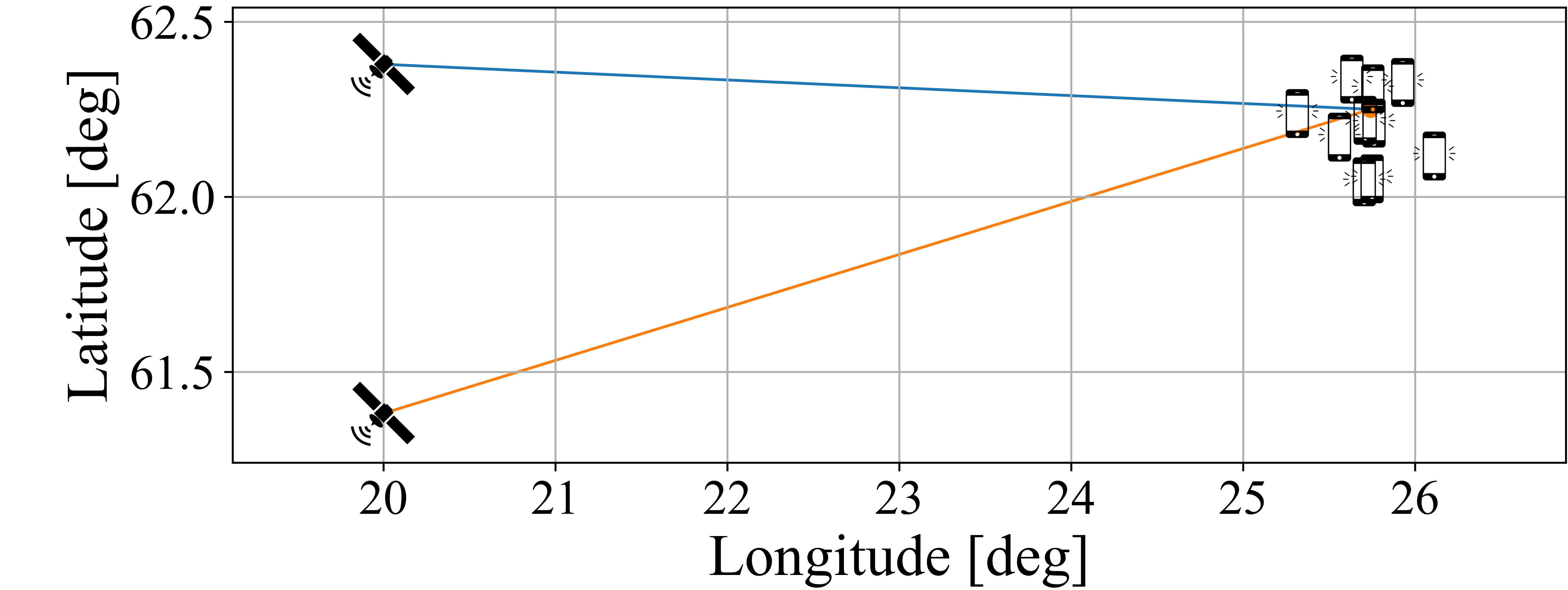}
    \caption{Simulation scenario at the beginning of a simulation. The lines depict the relations of the satellites toward the target points of the beams.}
    \label{fig:scenario}
\end{figure}

In the considered scenario (see Fig.~\ref{fig:scenario}), there are two transparent payload satellites, that is, the gNBs are on the ground and the satellites repeat the Nr-U signal. The Xn delay is considered 2 ms. Out-of-order delivery in PDCP is considered. Each satellite has its center beam directed to the same target point, operating on separate frequencies to avoid interference. Only the center beams are considered for statistics collection. There are two additional layers of wraparound beams for each of the satellites that are used to introduce interference. There is one full buffer UE in each of the wraparound beams. Frequency Reuse Factor (FRF) 3 is considered. The satellites fly from east to west in longitudinal orbits around 7.56 km/s at 600 km orbit \cite{38821}.

{}

Ten UEs are uniformly placed around the target point of the center beams. These UEs use cell selection to connect to the strongest cell and require reliable transmission with low tolerance for errors. The UEs require Constant Bit Rate (CBR) traffic. Only user link is considered. MC is activated for a UE when the RSRP of a neighbor cell plus an offset reaches the level of the serving cell, that is, A3 event (different MC activation schemes can be utilized based on specific requirements and needs). With the chosen offset (10 dBm), all the UEs requiring high reliability get MC activated. The UEs have Doppler mobility, simulating speed without actual movement, which is used for correlating channel updates, such as shadowing. Actual user mobility has not been implemented.

The channel condition from LOS to NLOS and vice versa changes dynamically based on the position of the serving satellite in relation to a UE. If the change in position exceeds a cube with side lengths of 3.5 km (chosen experimentally to account for changes to the LOS/NLOS conditions), a new LOS condition for the channel is randomly chosen based on \cite[Table 6.6.1-1]{38811}. For example, in a rural scenario with an elevation angle of 60°, the new LOS probability is 94\%. While more complex models may be useful in the future, the current model is sufficient to evaluate the benefits of PD, and the conclusions hold with more complicated models.

Simulations are run with three configurations: PD off, blind PD, and PD based on HARQ feedback. Each configuration is run 80 times with different RNG seeds, leading to variability in UE positions. Cumulative Distribution Function (CDF) statistics are combined from different RNG runs, while scalar statistics are averaged. The simulation time is 10 s with a warmup time of 0.5 s during which statistics are not collected. For PD based on HARQ feedback, the duplication duration after receiving NACK as HARQ feedback is 50 ms. In general, it should be set such that it captures consecutive NACKs, that is, at least the roundtrip time and processing delays. The simulation parameters are listed in Table~\ref{table:params}.

\begin{table}[]
\caption{Simulation parameters.}
\label{table:params}
\begin{tabularx}{\linewidth}{l|L}
\hline
\textbf{Parameter}               & \textbf{Value}                                   \\ \hline
Simulation time & 10.0 s                                                         \\
Warmup time & 0.5 s                                                              \\
Satellite payload & Transparent                                                              \\
Satellite mobility & Moving                                                      \\
UE mobility & Doppler (3 km/h)                                                   \\
Beam deployment & Quasi-Earth Fixed                                              \\
Satellite starting positions & Lat: 62.38°/61.38°, Lon: 20.00/20.00° \\
Center beam target points & Lat: 62.25°, Lon: 25.74°                 \\
NTN channel condition & Dynamic                                                  \\
UEs per center beam & 10                                                         \\
Wraparound & Two layers of beams with one full buffer UE in each                 \\
NTN scenario & Rural                                                             \\
Bandwidth per NTN beam & 10 Mhz                                                  \\
FRF & 3                                                                          \\
NTN carrier frequency & 2 GHz (S-band)                                           \\
Satellite orbit & LEO 600 km                                                     \\
Satellite parameter set & Set 1 \cite[Table 6.1.1.1-1]{38821}                      \\
UE antenna type & Omnidirectional                                                       \\
Traffic & CBR with UDP                                                           \\
UDP Packet Size & 32 B                                                           \\
UDP Packet Interval per UE & 20 ms                                               \\
Offset for SN addition & 10 dBm                                                  \\
\makecell{Duplication time after NACK\\received as HARQ feedback} & 50 ms                    \\
Xn delay & 2 ms                                                              \\
HARQ & Enabled with one retransmission                                           \\
Scheduler & Round Robin                                                          \\
RNG Runs & 80                                               \\ \hline
\end{tabularx}
\end{table}

\subsection{Results}

\begin{figure}[htb!]
    \centering
    \includegraphics[trim={0 0 0 0},clip, width=.6\linewidth]{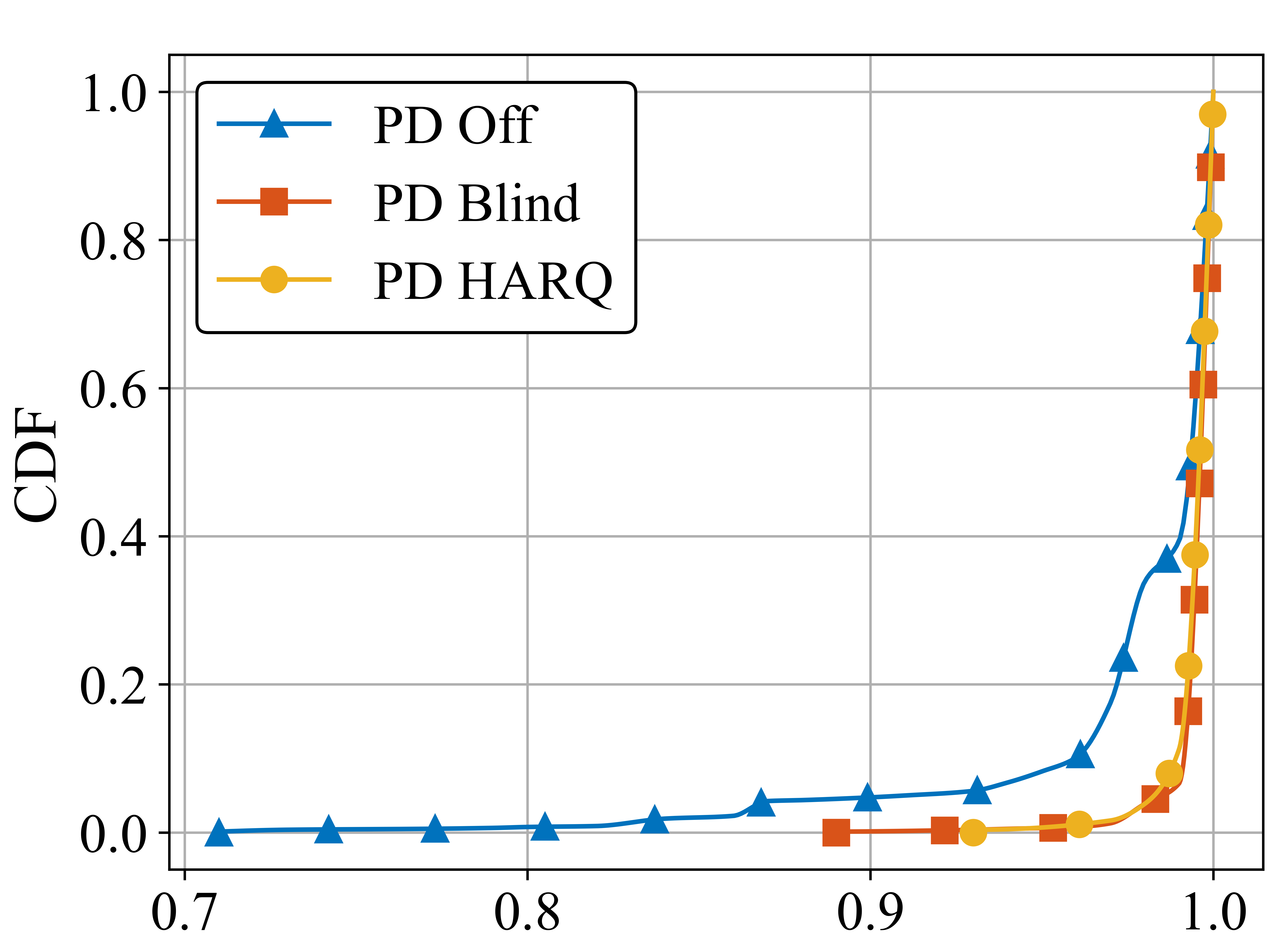}
    \caption{CDF of the users' application packet success rate.}
    \label{fig:success}
\end{figure}

Figure~\ref{fig:success} shows the CDF of the packet success rate for the users. For convenience, the statistics are summarized in Table~\ref{table:results}. When PD is not used, the mean success rate is 98.38\%, whereas when PD is used it is 99.86\% and 99.82\% respectively for blind and HARQ-based duplications. Both PD schemes enhance the success percentage by around 1.5 percentage points. Further, the HARQ-based duplication scheme performs marginally worse than the blind duplication scheme.

{}

\begin{figure}[htb!]
    \centering
    \includegraphics[trim={0 0 0 0},clip, width=.6\linewidth]{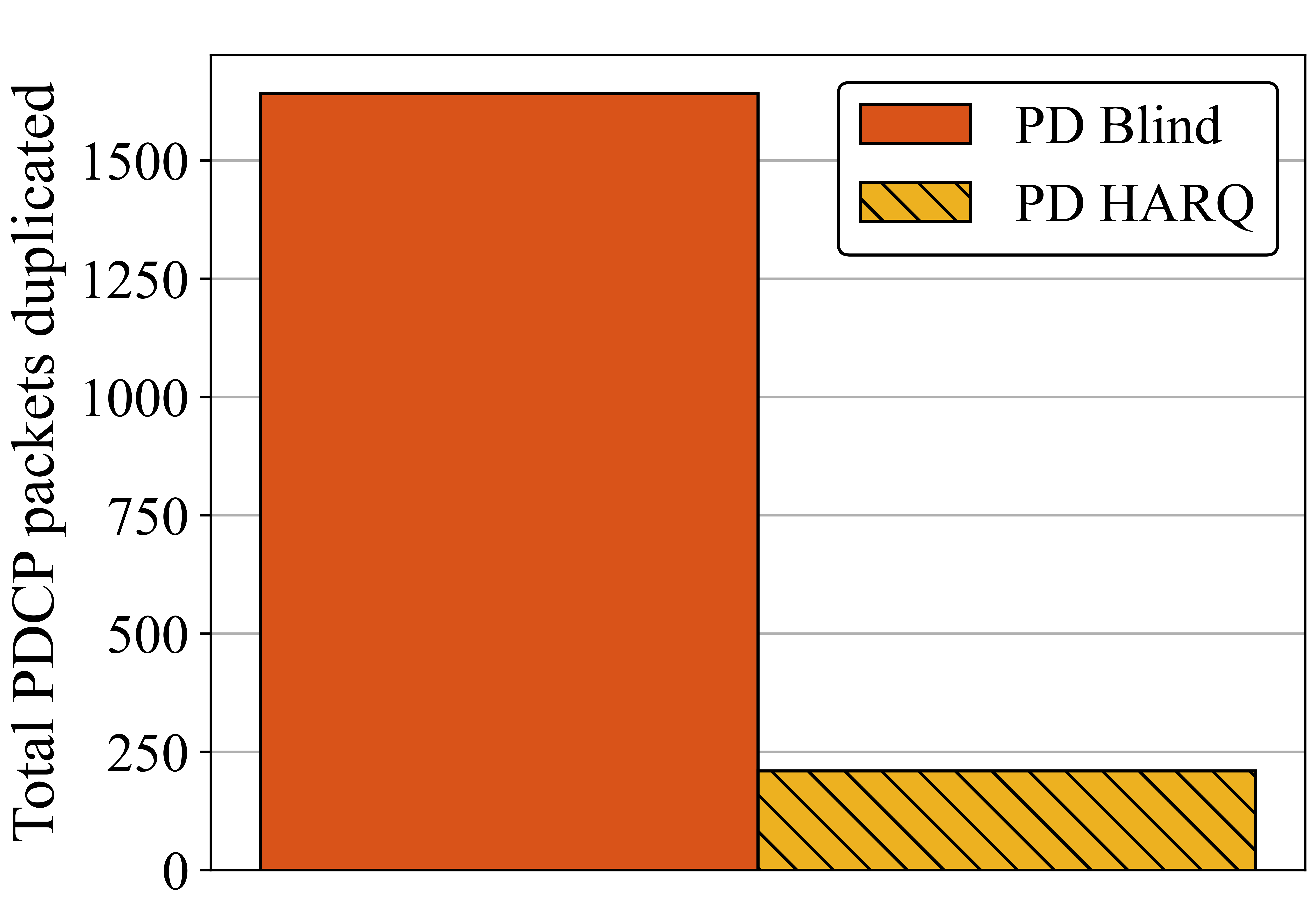}
    \caption{Total PDCP packets duplicated.}
    \label{fig:duplicates}
\end{figure}

Figure~\ref{fig:duplicates}. shows the total number of PDCP packets duplicated (i.e., for all the users, averaged over the RNG runs) for the different duplication schemes. The number of PDCP duplicates for the HARQ-based scheme is 12.78\% of the duplicates in the blind scheme.

{}

Both PD schemes significantly improve reliability, with the blind scheme performing slightly better than the HARQ-based scheme. However, the blind scheme results in more wasted network resources due to the duplication of all PDCP packets and the subsequent dropping of duplicate packets by the UE.

\begin{table}[]
\caption{Simulation results.}
\label{table:results}
\begin{tabularx}{\linewidth}{l|l|l|l}
\hline
\thead{\textbf{Simulation}} & \thead{\textbf{Mean success}\\\textbf{rate {[}\%{]}}} & \thead{\textbf{5th percentile}\\\textbf{ success rate {[}\%{]}}} & \thead{\textbf{PDCP packets}\\\textbf{duplicated}} 

\\ \hline

PD off     & 98.38                      & 91.00                        & -                    
\\
PD blind   & 99.86                      & 98.41                      & 1641                          
\\
PD HARQ    & 99.82                      & 98.15                      & 209.8                       
\\ \hline

\end{tabularx}
\end{table}

\section{Conclusion}
\label{sec:conclusions}

In this paper, reliability through PD in NTNs was researched. Through PD, reliability can be increased by sending the same data over different paths. To decrease the number of excess duplicates, a dynamic PD scheme was proposed. By system-level simulations, the proposed scheme was evaluated in a scenario with two transparent payload LEO satellites. The number of duplicated PDCP packets was reduced by 87.22\% compared to blind duplication while only marginally affecting the reliability.

In the future, PD activation could be enhanced by considering multiple HARQ feedbacks or error rates over a period of time. Currently, packets below the PDCP layer cannot be duplicated after PD activation, but this limitation could be overcome through cross-layer designs, such as buffering packets at the PDCP layer until feedback is received. Additionally, while the 99.9\% reliability requirement of HRC-s was nearly met, further work is needed to closely address constellation and scenario design to fully achieve this requirement. 

\section*{Acknowledgment}
This work has been partially funded by the European Union Horizon-2020 Project DYNASAT (Dynamic Spectrum Sharing and Bandwidth-Efficient Techniques for High-Throughput MIMO Satellite Systems) under Grant Agreement 101004145. The views expressed are those of the authors and do not necessarily represent the project. The Commission is not liable for any use that may be made of any of the information contained therein.

\vspace{6pt}
\bibliography{references} 

\begin{thebibliography}{10}
\providecommand{\url}[1]{#1}
\csname url@samestyle\endcsname
\providecommand{\newblock}{\relax}
\providecommand{\bibinfo}[2]{#2}
\providecommand{\BIBentrySTDinterwordspacing}{\spaceskip=0pt\relax}
\providecommand{\BIBentryALTinterwordstretchfactor}{4}
\providecommand{\BIBentryALTinterwordspacing}{\spaceskip=\fontdimen2\font plus
\BIBentryALTinterwordstretchfactor\fontdimen3\font minus
  \fontdimen4\font\relax}
\providecommand{\BIBforeignlanguage}[2]{{%
\expandafter\ifx\csname l@#1\endcsname\relax
\typeout{** WARNING: IEEEtran.bst: No hyphenation pattern has been}%
\typeout{** loaded for the language `#1'. Using the pattern for}%
\typeout{** the default language instead.}%
\else
\language=\csname l@#1\endcsname
\fi
#2}}
\providecommand{\BIBdecl}{\relax}
\BIBdecl

\bibitem{wid}
``{New WI}: {NR NTN} (non-terrestrial networks) enhancements,'' 3GPP,
  RP-213690, Dec. 2021.

\bibitem{itu}
``Vision, requirements and evaluation guidelines for satellite radio
  interface(s) of {IMT}-2020,'' ITU, M.2514-0, Sept. 2022.

\bibitem{37340}
``{TS} 37.340: {NR}; multi-connectivity; overall description; stage-2,'' 3GPP,
  V16.7.0, Sept. 2021.

\bibitem{8884124}
M.-T. Suer, C.~Thein, H.~Tchouankem, and L.~Wolf, ``Multi-connectivity as an
  enabler for reliable low latency communications—an overview,'' \emph{IEEE
  Communications Surveys Tutorials}, vol.~22, no.~1, pp. 156--169, 2020.

\bibitem{8491093}
N.~H. Mahmood, M.~Lopez, D.~Laselva, K.~Pedersen, and G.~Berardinelli,
  ``Reliability oriented dual connectivity for {URLLC} services in {5G} {New}
  {Radio},'' \emph{2018 15th International Symposium on Wireless Communication
  Systems (ISWCS)}, pp. 1--6, 2018.

\bibitem{9027289}
S.~M. Rayavarapu, S.~D. Amuru, and K.~Kiran, ``Dynamic control of packet
  duplication in {5G-NR} dual connectivity architecture,'' in \emph{2020
  International Conference on COMmunication Systems \& NETworkS (COMSNETS)},
  2020, pp. 539--542.

\bibitem{8377054}
J.~Rao and S.~Vrzic, ``Packet duplication for {URLLC} in {5G} dual connectivity
  architecture,'' in \emph{2018 IEEE Wireless Communications and Networking
  Conference (WCNC)}, 2018, pp. 1--6.

\bibitem{9128536}
M.~Centenaro, D.~Laselva, J.~Steiner, K.~Pedersen, and P.~Mogensen,
  ``Resource-efficient dual connectivity for ultra-reliable low-latency
  communication,'' in \emph{2020 IEEE 91st Vehicular Technology Conference
  (VTC2020-Spring)}, 2020, pp. 1--5.

\bibitem{10008752}
M.~López, S.~B. Damsgaard, I.~Rodríguez, and P.~Mogensen, ``An empirical
  analysis of multi-connectivity between {5G} terrestrial and {LEO} satellite
  networks,'' in \emph{2022 IEEE Globecom Workshops (GC Wkshps)}, 2022, pp.
  1115--1120.

\bibitem{majamaa2022multiconnectivity}
M.~Majamaa, H.~Martikainen, L.~Sormunen, and J.~Puttonen, ``Multi-connectivity
  in {5G} and beyond non-terrestrial networks,'' \emph{Journal of
  Communications Software and Systems}, vol.~18, no.~4, pp. 350--358, 12 2022.

\bibitem{38323}
``{TS 38.323: 5G; NR;} packet data convergence protocol ({PDCP})
  specification,'' 3GPP, V16.2.0, Nov. 2020.

\bibitem{38331}
``{TS 38.331: 5G; NR;} radio resource control ({RRC}); protocol
  specification,'' 3GPP, V16.1.0, July 2020.

\bibitem{ntn}
J.~Puttonen, L.~Sormunen, H.~Martikainen, S.~Rantanen, and J.~Kurjenniemi, ``A
  system simulator for {5G} non-terrestrial network evaluations,'' \emph{2021
  IEEE 22nd International Symposium on a World of Wireless, Mobile and
  Multimedia Networks (WoWMoM)}, pp. 292--297, 2021.

\bibitem{Riley2010}
G.~F. Riley and T.~R. Henderson, ``The ns-3 network simulator,'' in
  \emph{Modeling and Tools for Network Simulation}, K.~Wehrle,
  M.~G{\"u}ne{\c{s}}, and J.~Gross, Eds.\hskip 1em plus 0.5em minus 0.4em\relax
  Berlin, Heidelberg: Springer Berlin Heidelberg, 2010, pp. 15--34.

\bibitem{Patriciello2019AnES}
N.~Patriciello, S.~Lag{\'e}n, B.~Bojovi{\'c}, and L.~Giupponi, ``An {E2E}
  simulator for {5G} {NR} networks,'' \emph{Simulation Modelling Practice and
  Theory}, vol.~96, 2019.

\bibitem{lenalte}
N.~Baldo, M.~Miozzo, M.~Requena-Esteso, and J.~Nin-Guerrero, ``An open source
  product-oriented {LTE} network simulator based on ns-3,'' in
  \emph{Proceedings of the 14th ACM International Conference on Modeling,
  Analysis and Simulation of Wireless and Mobile Systems}, ser. MSWiM
  '11.\hskip 1em plus 0.5em minus 0.4em\relax New York, NY, USA: Association
  for Computing Machinery, 2011, p. 293–298.

\bibitem{38821}
``{TR} 38.821: Solutions for {NR} to support non-terrestrial networks
  ({NTN}),'' 3GPP, V16.0.0, Jan. 2020.

\bibitem{38811}
``{TR} 38.811: Study on new radio ({NR}) to support non-terrestrial networks,''
  3GPP, V15.4.0, Sept. 2020.

\bibitem{9149179}
A.~Guidotti, A.~Vanelli-Coralli, A.~Mengali, and S.~Cioni, ``Non-terrestrial
  networks: Link budget analysis,'' in \emph{ICC 2020 - 2020 IEEE International
  Conference on Communications (ICC)}, 2020, pp. 1--7.

\end{thebibliography}
\bibliographystyle{IEEEtran}

\end{document}